\documentclass{aa}

\usepackage{graphicx}
\usepackage{subcaption}
\usepackage{layouts}
\usepackage{ amssymb }
\usepackage{mathrsfs}

\usepackage{txfonts}
\usepackage{xcolor}
\usepackage{hyperref}
\begin{document}

   \title{Detection of coherent low-frequency radio bursts from weak-line T Tauri stars}

   %\subtitle{}

   \author{A. Feeney-Johansson
          \inst{1,2}
          \and
          S. J. D. Purser
          \inst{1}
          \and
          T. P. Ray
          \inst{1}
          \and
          A. A. Vidotto
          \inst{2}
          \and
          J. Eisl\"offel
          \inst{3}
          \and
          J. R. Callingham
          \inst{4,5}
          \and
          T. W. Shimwell
          \inst{5,4}
          \and
          H. K. Vedantham
          \inst{6}
          \and
          G. Hallinan
          \inst{7}
          \and
          C. Tasse
          \inst{8,9}
          }

   \institute{Dublin Institute for Advanced Studies, Astronomy \& Astrophysics Section, 31 Fitzwilliam Place, Dublin, D02 XF86, Ireland\\
              \email{antonfj@cp.dias.ie}
        \and
            School of Physics, Trinity College Dublin, College Green, Dublin 2, Ireland
        \and
            Th\"uringer Landessternwarte Tautenburg, Sternwarte 5, D-07778, Tautenburg, Germany
        \and
        Leiden Observatory, Leiden University, PO Box 9513, 2300 RA, Leiden, The Netherlands
        \and
        ASTRON, Netherlands Institute for Radio Astronomy, 
        Oude Hoogeveensedijk 4, Dwingeloo, 7991 PD, The Netherlands
        \and
        Kapteyn Astronomical Institute, University of Groningen, 
        PO Box 72, 97200 AB, Groningen, The Netherlands
        \and
        Cahill Center for Astronomy and Astrophysics, California Institute of Technology, Pasadena, CA 91125, USA
        \and
        GEPI \& USN, Observatoire de Paris, CNRS, Universit\'e Paris Diderot, 5 place Jules Janssen, 92190 Meudon, France 
        \and
        Centre for Radio Astronomy Techniques and Technologies, Department of Physics and Electronics, Rhodes University, Grahamstown 6140, South Africa
        }

   \date{}

% \abstract{}{}{}{}{} 
% 5 {} token are mandatory
 
  %\abstract
   %{textwidth in cm: \printinunitsof{cm}\prntlen{\textwidth}; textheight in cm: \printinunitsof{cm}\prntlen{\textheight}\\
%textwidth in inches: \printinunitsof{in}\prntlen{\textwidth}; textheight in inches: \printinunitsof{in}\prntlen{\textheight}\\
%columnwidth in cm: \printinunitsof{cm}\prntlen{\columnwidth}; columnwidth in inches: \printinunitsof{in}\prntlen{\columnwidth}}

    \abstract{In recent years, thanks to new facilities such as LOFAR capable of sensitive observations, much work has been done on the detection of stellar radio emission at low frequencies. Such emission has commonly been shown to be coherent emission, generally attributed to electron-cyclotron maser emission, and has usually been detected from main-sequence M dwarfs. Here we report the first detection of coherent emission at low frequencies from T Tauri stars, which are known to be associated with high levels of stellar activity. Using LOFAR, we have detected several bright radio bursts at 150 MHz from two weak-line T Tauri stars: KPNO-Tau 14 and LkCa 4. All of the bursts have high brightness temperatures ($10^{13} - 10^{14}\ \mathrm{K}$) and high circular polarization fractions (60 - 90 \%), indicating that they must be due to a coherent emission mechanism. This could be either plasma emission or electron-cyclotron maser (ECM) emission. Due to the exceptionally high brightness temperatures seen in at least one of the bursts ($\geq 10^{14}\ \mathrm{K}$), and the high circular polarization levels, it seems unlikely that plasma emission could be the source and so ECM is favoured as the most likely emission mechanism. Assuming this is the case, the required magnetic field in the emission regions would be 40 - 70 G. We determine that the most likely method of generating ECM emission is plasma co-rotation breakdown in the stellar magnetosphere. There remains the possibility, however, it could be due to an interaction with an orbiting exoplanet.}


   \maketitle
%
%-------------------------------------------------------------------

\begin{table*}[t]
    \caption{Primary characteristics of KPNO-Tau 14 and LkCa 4.}
    \centering
    \begin{tabular}{c c c}
        \hline
         & KPNO-Tau 14 & LkCa 4 \\
        \hline
        Spectral Type & M6 & K7 \\
        Age (Myrs) & 1 \textsuperscript{1} & 2 \textsuperscript{6} / 0.5 \textsuperscript{7}\\
        Distance (pc) & $151.9^{+5.1}_{-4.8}$ \textsuperscript{2} & $129.4\pm0.6$ \textsuperscript{2} \\
        Mass (M$_{\odot}$) & 0.1\textsuperscript{1} &  $0.79\pm0.05$ \textsuperscript{6} / 0.15 - 0.30 \textsuperscript{7}\\
        Radius (R$_{\odot}$) & $\sim1$  \textsuperscript{3} & 2.3 \textsuperscript{7}\\
        Rotation Period & $1.86\pm 0.01$ \textsuperscript{4} & $3.37 \pm 0.01$ \textsuperscript{8} \\
        Surface Magnetic Field Strength (kG) & - & 1.6 \textsuperscript{6} \\
        Coronal Temperature (MK) & 12.6 \textsuperscript{5} & - \\
        \hline
    \end{tabular}
    \tablebib{
    (1) \citet{Luhman2003}; (2) \citet{Bailer-Jones2018}; (3) Derived from \citet{Baraffe2015} models; (4) \citet{Scholz2018}; (5) \citet{Gudel2007}; (6) \citet{Donati2014}; (7) \citet{Gully-Santiago2017}; (8) \citet{Grankin2008}
    }
    \tablefoot{Due to the uncertainty in the age and mass of LkCa4, two values are given for these parameters.}
    \label{PrimaryCharacteristics}
\end{table*}

\section{Introduction}

% Talk about typical thermal radio emission from classical and weak-lined T Tauri stars 

T Tauri stars (TTSs) are low-mass ($<2~M_{\odot}$) young stellar objects (YSOs) that have evolved to the state where they are optically visible. They are frequently associated with magnetic activity seen in the form of flaring and magnetospheric accretion bursts \citep{Feigelson1999}. TTSs can be divided into two categories: classical T Tauri stars (CTTSs), which are surrounded by a circumstellar accretion disk, and weak-line T Tauri stars (WTTSs), where most, if not all, of the disk has dissipated. Flares from these objects are thought to be caused by magnetic reconnection events, similar to the ones observed on the Sun but with energies up to $10^6$ times larger \citep[e.g.,][]{Fernandez2004}. In addition, while reconnection in the Sun and other main sequence stars occurs in magnetic fields anchored in the star, reconnection in YSOs can occur in magnetic fields connecting the star and the disk \citep{Waterfall2019}.

Flaring of WTTSs has been observed at a variety of wavelengths from the X-ray \citep[e.g.,][]{Uzawa2011} to radio regimes \citep[e.g.,][]{Forbrich2017}. In the radio band, observations have normally been made at cm-wavelengths ($\geq 1\ \mathrm{GHz}$) where any detected emission is generally attributed to incoherent gyro-synchrotron emission from energetic electrons excited in magnetic reconnection events \citep{Waterfall2019}.

Among main sequence stars, radio bursts have primarily been observed from M dwarfs, which are known to be magnetically active \citep{Villadsen2019}. At frequencies below or around 1 GHz, such bursts have been found to be coherent emission, generated by either plasma emission or electron-cyclotron maser (ECM) emission. In both cases coherent radio emission is generally characterised by very high brightness temperatures \citep{Dulk1985}.

For example, one well known flare star, UV Ceti, was detected at 154~MHz using the Murchison Widefield Array \citep{Lynch2017a}.
This emission was determined to be coherent based on its high brightness temperature ($10^{13} - 10^{14}\ \mathrm{K}$) and, most likely, ECM emission due to the presence of elliptical polarisation, which is thought to only be generated in the x-mode \citep{Dulk1994}. From theory, it is predicted that ECM emission is entirely polarised in the x-mode while plasma emission is polarised in the o-mode \citep{Dulk1985}.

In addition, \citet{Zic2019} detected several periodic, coherent bursts from UV Ceti. The period of the bursts was consistent with the rotation period of the star and based on the presence of high brightness temperatures and elliptical polarization, the bursts were determined to be strongly-beamed ECM emission originating from a low-density cavity within the magnetosphere of UV Ceti. This illustrates that auroral emission is possible in main sequence M dwarfs, similar to that seen in brown dwarfs and Solar System planets \citep{Pineda2017,Zarka1998}.

Recently, coherent radio emission has also been detected from the flare star CR Draconis by \citet{Callingham2021} using LOFAR \citep{VanHaarlem2013}. When observed for a total of 6.5 days over one year, three bright radio bursts were detected in addition to an almost steady radio flux. Due to the high brightness temperatures, it was determined that the emission must be coherent. Moreover, it shows high levels of left-handed circular polarisation at all epochs. The consistent handedness of the circular polarisation indicates that the emission originates from a region with the same magnetic polarity. Thus it is unlikely to be due to plasma emission generated by flares since the polarity would be expected to flip due to flares occurring across the stellar disk. In addition, elliptical polarisation was also seen in one of the bursts detected. Based on these factors, it was determined that the emission must be ECM emission.

However, very little work has been done on searching for radio emission from YSOs at low ($ <1   \rm{GHz}$) radio frequencies.  The detection of such emission can give us important information on the presence of any high energy particles and magnetic fields in YSO outflows \citep{Ainsworth2014, Feeney-Johansson2019}, provide evidence for possible star-planet interactions \citep{Vedantham2020, KavanaghSubmitted}, and, help evaluate the importance of Coronal Mass Ejection (CME) events in dissipating proto-planetary disks \citep{Schnepf2015}.

While in its infancy, the study of YSOs at low radio frequencies is nevertheless a growing field thanks to new facilities such as LOFAR.  Here one can utilise not only single target observations but also surveys such as the LOFAR Two-metre Sky Survey (LoTSS) that allow steady and transient sources to be found \citep{Shimwell2017}.

With these ideas in mind, we have started an initial study using the High-Band Array (HBA) of LOFAR of the Taurus Molecular Cloud (TMC) centred on approximately 150~MHz. Note that the HBA was used for this study to take advantage of its much higher sensitivity compared to the Low-Band Array (LBA) of LOFAR. Here we discuss two weak-lined T Tauri stars in the TMC, KPNO-Tau 14 and LkCa 4, the primary characteristics of which are listed in Table \ref{PrimaryCharacteristics}.

KPNO-Tau~14 is a very-low-mass ($0.1 {\rm M}_{\sun}$) WTTS of spectral type M6 \citep{Luhman2003}. This puts the star very close to the brown dwarf upper limit ($< 0.08 {\rm M}_{\sun}$). Based on \textit{Gaia} measurements \citep{Gaia2018}, its distance is $151.9^{+5.1}_{-4.8}\ \mathrm{pc}$ \citep{Bailer-Jones2018}. No significant level of accretion has been detected in this YSO and there is no infrared excess in its spectrum \citep{Muzerolle2005, Mohanty2005}, suggesting that there is no dusty disk present. \citet{Mohanty2005} found evidence that KPNO-Tau~14 may be a spectroscopic binary, based on the measurement of a double peak in the cross-correlation profile of its H$\alpha$ line. However, a high-resolution imaging survey by \citet{Kraus2006} using HST was unable to resolve the system, giving an upper limit for the projected binary separation of $<$4~au. Magnetic activity has been inferred from KPNO-Tau~14. More precisely X-ray emission was detected from this star in the XMM-Newton extended survey of the Taurus Molecular Cloud (XEST) by  \citet{Gudel2007}. During these observations, an X-ray flare was found with a luminosity of $1.58 \times 10^{30}\ \mathrm{erg\ s^{-1}}$ (corrected for extinction) and a duration of $>6000\ \mathrm{s}$ \citep{Stelzer2007}. For comparison, the median X-ray luminosity for T Tauri stars detected by \citet{Gudel2007} was $0.7 \times 10^{30}\ \mathrm{erg\ s^{-1}}$.

LkCa~4 is another WTTS of spectral type K7. Based on \textit{Gaia}, the distance to this star is $129.4\pm0.6\ \mathrm{pc}$ \citep{Bailer-Jones2018}. Like KPNO-Tau~14 it has no infrared excess, indicating that its dusty disk has dissipated or at least that significant grain growth has occurred reducing its infrared emission \citep{Esplin2014}.
LkCa~4 is known to be magnetically very active. This is deduced from its high level of starspot coverage and intense flaring. In fact,  \citet{Gully-Santiago2017} discovered, using model atmospheres, that its spectrum is best fitted with two temperature components: a hot photosphere of $T_{hot} \sim 4100\ \mathrm{K}$ covering 20\% of the surface and a cool component with $T_{cool} \sim 2700-3000\ \mathrm{K}$ accounting for the remainder. This suggests a remarkable starspot coverage of 80\%. Flaring activity has previously been seen at radio frequencies by \citet{Dzib2015}, who observed mildly circularly polarised emission at 4.5 GHz and 7.5 GHz. The emission was highly variable in Stokes I over 3 epochs, with changes of $92.7\pm22.7$\% at 4.5 GHz and $90.8\pm20.1$\% at 7.5 GHz. The circular polarisation fractions measured were 9.6\% at 4.5 GHz and 18.8\% at 7.5 GHz. LkCa~4 has also been detected in X-rays by Chandra with a measured luminosity corrected for extinction of $2.0 \times 10^{30}\ \mathrm{erg\ s^{-1}}$ \citep{Yang2012}.

\begin{table*}[t]
    \caption{Details of the LOFAR observations of KPNO-Tau 14.}
    \centering
    \begin{tabular}{c c c c c c c c}
    \hline 
    Project code & LoTSS pointing & Date & Time & Duration & $\nu$ & $\Delta \nu$ & $\sigma_{\mathrm{rms}}$ \\ 
     & & & & (hrs) & (MHz) & (MHz) & ($\mu$Jy) \\
    \hline 
    LC1\_001 & - & 2013-11-30 & 19:15:00 & 8.0 & 152 & 74 & 210 \\
    L12\_015 & P068+26 & 2019-06-28 & 05:11:00 & 8.0 & 144 & 48 & 130 \\
    \hline 
    \end{tabular}
    \tablefoot{The Date and Time columns refer to the time in UTC at the start of each observation. $\nu$ is the central frequency of the observation and $\Delta \nu$ is the bandwidth. $\sigma_{\mathrm{rms}}$ is the root-mean-square noise in Stokes I at the position of KPNO-Tau 14.}
    \label{KPNO-Tau_14_observations}
\end{table*}

\begin{table*}[t]
    \caption{Details of the LOFAR observations of LkCa 4.}
    \centering
    \begin{tabular}{c c c c c c c c c}
    \hline 
    Project code & LoTSS pointing & Date & Time & Duration & $\nu$ & $\Delta \nu$ & $\sigma_{\mathrm{rms}}$ & Phase \\
     & & & & (hrs) & (MHz) & (MHz) & ($\mu$Jy) &  \\
    \hline 
    L12\_015 & P061+29 & 2019-06-28 & 05:11:00 & 8.0 & 144 & 48 &  230 & $0.00 - 0.10$ \\
    L12\_015 & P062+26 & 2019-07-06 & 04:52:15 & 8.0 & 144 & 48 &  240 & $0.37 - 0.47$ \\
    L12\_015 & P065+26 & 2019-07-07 & 04:59:59 & 8.0 & 144 & 48 &  280 & $0.67 - 0.77$ \\
    L12\_015 & P064+29 & 2019-08-10 & 02:11:00 & 8.0 & 144 & 48 &  150 & $0.71 - 0.81$\\ 
    \hline 
    \end{tabular} 
    \label{LkCa_4_observations}
    \tablefoot{The Date and Time columns refer to the time in UTC at the start of each observation. $\nu$ is the central frequency of the observation and $\Delta \nu$ is the bandwidth. $\sigma_{\mathrm{rms}}$ is the root-mean-square noise in Stokes I at the position of LkCa 4. The rotation phase is calculated using the known rotation period of LkCa 4 of 3.37 d \citep{Grankin2008} such that phase 0.0 corresponds to the beginning of the observation of LoTSS pointing P061+29 (2019-06-28 05:11:00).}
\end{table*}

The magnetic field of LkCa~4 was modelled by \citet{Donati2014} who found that it possesses a strong and mainly axisymmetric magnetic field with a poloidal component of $\approx$ 2 kG and a toroidal component of $\approx$ 1 kG.
% Maybe need to write about orientation of field. Although it might be better to put that later in results when comparing the circular polarisation orientation with the field orientation.
There is a large degree of uncertainty regarding the mass and age of LkCa~4, due to its high level of starspot coverage. While previous estimates, using an effective temperature of 4000\,K, had found mass and age estimates of $0.79\pm0.05\ \mathrm{M_{\odot}}$ and 2 Myr, \citet{Gully-Santiago2017} find an effective temperature of 3180 K based on their two-temperature component model. Comparing this to stellar evolutionary models \citep{Baraffe2015} gives mass and age estimates of $0.15-0.30\ \mathrm{M_{\odot}}$ and $\sim0.5\ \mathrm{Myr}$ respectively. These evolutionary models assume the star to be unspotted, however, and thus there remain major uncertainties.

In this paper, we report the serendipitous detection of bright radio bursts at low frequencies from these two sources using LOFAR. In section 2, we describe how the observations were carried out and the data reduction process. In section 3, we show the resulting images and light curves of the observations that were obtained and describe the characteristics of the bursts that were detected. In section 4, we then discuss which of the possible emission mechanisms can explain the observations. In section 5, we discuss whether the ECM mechanism could also be observed from YSOs at higher radio frequencies. Finally, in section 6, we present our concluding remarks.

\section{Observations and Data Reduction}

\subsection{2013 LOFAR Observations}
As part of a project to search for emission from the jets of low-mass YSOs (LOFAR project code: LC1\_\_001), the classical T Tauri star DG Tau A, located in the TMC, was observed using the HBA on 2013 November 30 - December 1 (epoch 2013.91). The total observing time was 8 hours with a central frequency of 152~MHz and a bandwidth of 74~MHz \citep{Feeney-Johansson2019}. Details of this observation are given in Table \ref{KPNO-Tau_14_observations}

Due to the ionosphere and imperfect knowledge of beam shapes, there are important direction-dependent effects (DDEs) present in LOFAR data, that must be removed. Direction-independent calibration was first performed using the LOFAR calibration pipeline \textsc{Prefactor}\footnote{\url{https://github.com/lofar-astron/prefactor}} described by \citet{DeGasperin2019}. Subsequent direction-dependent calibration was carried out using the facet calibration pipeline \textsc{Factor}\footnote{\url{https://github.com/lofar-astron/factor}} described by \citet{VanWeeren2016}, followed by imaging using WSClean \citep{Offringa2014}. The resolution achieved in the final image was $5.88\arcsec \times 5.15\arcsec$. Full details of the process are given in \citet{Feeney-Johansson2019}.

Due to inaccuracies in LOFAR beam models \citep{VanWeeren2016, Shimwell2019}, the derived flux density scale is often not accurate. To account for this, in \citet{Feeney-Johansson2019} the integrated flux densities of several compact bright sources in the field were compared with their values in the TIFR GMRT Sky Survey \citep[TGSS,][]{Intema2017}. It was found that the average ratio of TGSS flux densities to LOFAR flux densities was $0.83\pm0.17$. However, there are known to be significant uncertainties with the flux density scale of TGSS. For this reason, and also to be consistent with the 2019 observations in the next section, the correction factor for the flux density scale was rederived using the NRAO VLA Sky Survey \citep[NVSS,][]{Condon1998} and the 6C catalogue \citep{Hales1988}. This technique is described in more detail in \citet{Hardcastle2021}. The flux-scale correction derived was $1.04\pm0.21$ and so the flux densities measured in the LOFAR image were corrected by this factor. \citet{Shimwell2019} found that there was still, conservatively, a $20\%$ error in the flux density scale of LOFAR images after correction. Therefore, there is estimated to be an error of $\pm0.21$ in the correction factor.

\begin{figure*}
    \includegraphics[width=\textwidth]{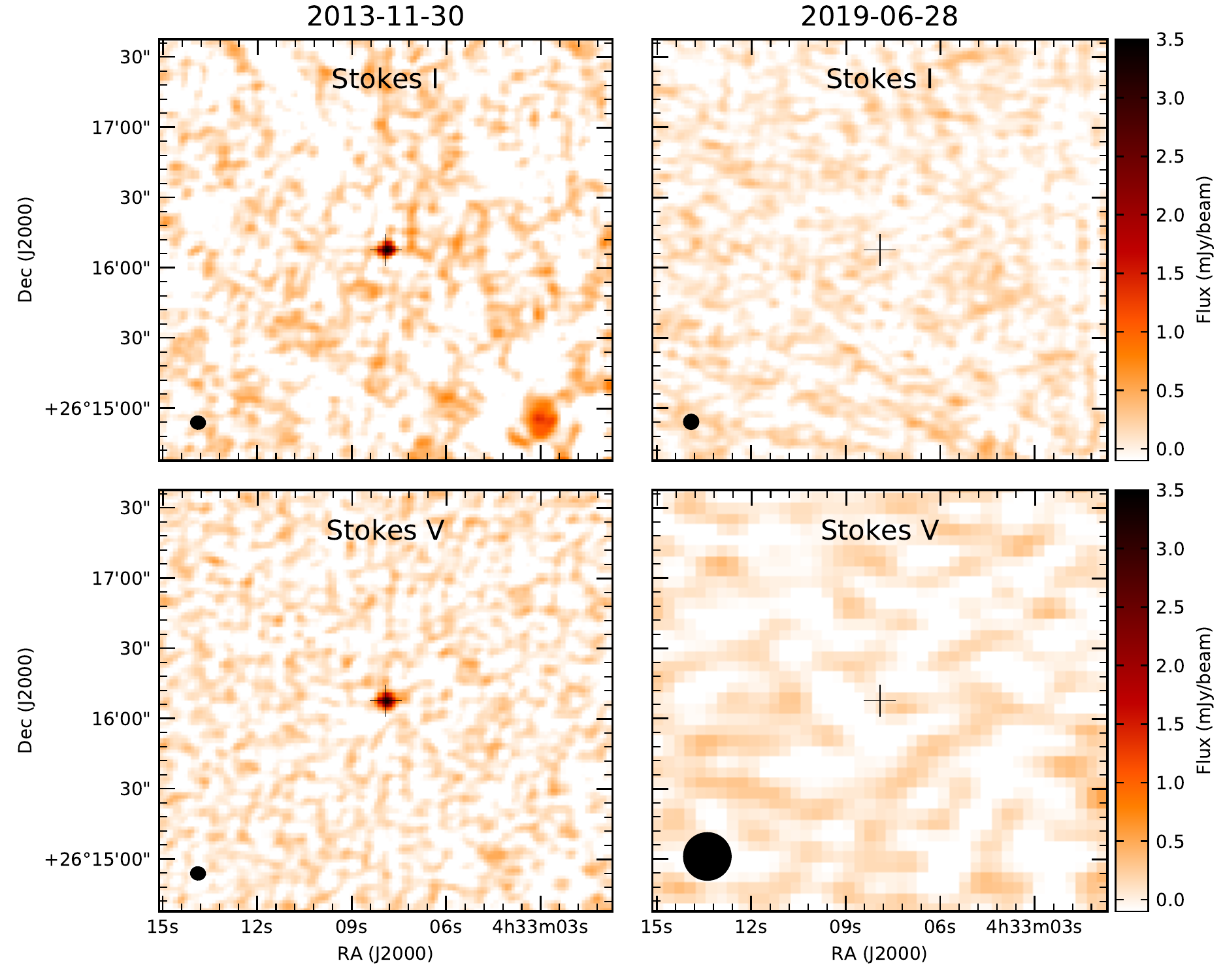}
    \caption{Stokes I (top) and Stokes V (bottom) images of KPNO-Tau~14 obtained from LOFAR at at two epochs: 2013 Nov 30 (left) and 2019 June 28 (right). The restoring beam used for the 2013 images was $5.88\arcsec \times 5.15\arcsec$. The restoring beams used for the 2019 LoTSS images were $6\arcsec \times 6\arcsec$ and $20\arcsec \times 20\arcsec$ for the Stokes I and Stokes V images respectively.  The optical position of KPNO-Tau 14 obtained from Gaia DR2 \citep{Gaia2018} is indicated by the cross.}
    \label{KPNO-Tau_14_images}
\end{figure*}

Although the main objective was to detect emission from the vicinity of DG~Tau A, the full field was examined for other YSOs. A catalogue of all radio sources down to a $5\sigma$ signal-to-noise threshold was extracted using PyBDSF \citep{Mohan2015} which was then compared with those of known YSOs in XEST \citep{Gudel2007}, the Gould Belt GHz VLA Survey  \citep{Dzib2015}, the Taurus Spitzer Survey \citep{Rebull2010}, and the Taurus WISE Survey \citep{Esplin2014}. 

Out of a total of 115 YSOs within the field, radio emission was detected at the position of 2 other YSOs in addition to DG~Tau~A: KPNO-Tau~14 and V1320~Tau, with the noise level ranging from 
$100\ \mu Jy\ \mathrm{beam}^{-1}$ 
at the centre of the field to 
$400\ \mu Jy\ \mathrm{beam}^{-1}$ 
near the edge of the field. While the flux of V1320~Tau is steady, KPNO-Tau~14 showed strong flaring during our observing run. As this paper focuses on radio bursts, V1320~Tau will be discussed later in a separate paper on low frequency emission from YSOs in the TMC (Feeney-Johansson et al., 2021, in preparation).

\subsection{2019 LOFAR Observations}
The LOFAR Two-Metre Sky Survey (LoTSS) is mapping the entire northern sky at 120-168~MHz with a target sensitivity of $\sim100\ \mathrm{\mu Jy\ beam^{-1}}$ \citep{Shimwell2017}. As part of this survey, the Taurus and Perseus Molecular Clouds were imaged in a co-observing proposal (LOFAR Project code:LC12\_\_015) in a series of observations during LOFAR Cycle 12, from June -- November 2019. These regions each contain at least 414 and 369 YSOs respectively  \citep{Esplin2014,Young2015}. Details of these observations are given in Tables \ref{KPNO-Tau_14_observations} and \ref{LkCa_4_observations}. Work on this survey is currently ongoing and the full results will be published in a future paper.

Direction-independent calibration was carried out on each of the observations using the \textsc{Prefactor} pipeline. To correct for DDEs, the data was processed using the DDF-Pipeline \citep{Shimwell2019,Tasse2020}, which uses the  KillMS package \citep{Smirnov2015} to solve for the DDEs and then uses DDFacet \citep{Tasse2018}, which applies the DDEs while imaging\footnote{See \url{https://github.com/saopicc} for the KillMS and DDFacet packages and \url{https://github.com/mhardcastle/ddf-pipeline} for the DDF-Pipeline}. The final output is a full-bandwidth Stokes I image of the field with a resolution of $6\arcsec$. In addition, a low resolution $20\arcsec$ Stokes V image  is also obtained.

As with the 2013 observations, and for each pointing, we extracted a list of sources for comparison with the previously cited YSO catalogues. In pointing P064+29, carried out on 2019 August 10, emission was detected towards LkCa~4. Similar to KPNO-Tau~14, this source also showed very strong flaring over the observing period. The position of LkCa~4 was also included in 3 other pointings: P061+29, P062+26, and P065+26, dated June 28, July 6, and July 7 2019 respectively. However, no emission was detected towards LkCa~4 in any of these earlier epochs. Likewise, the position of KPNO-Tau 14 was included in the pointing P068+26, dated 2019 June 28, but no emission was detected.

Similar to the 2013 LOFAR observations, the flux density scales for each of the LoTSS pointings were corrected due to errors in the beam models by deriving correction factors using NVSS and the 6C catalogue. For P064+29, the flux-scale correction derived was $1.24\pm0.25$ and so the flux density measurements in this pointing were multiplied by this factor. Again, based on \citet{Shimwell2019} there was assumed to be an error of $20\%$ in this correction factor.

\begin{figure}
    \centering
    \includegraphics[width=\hsize]{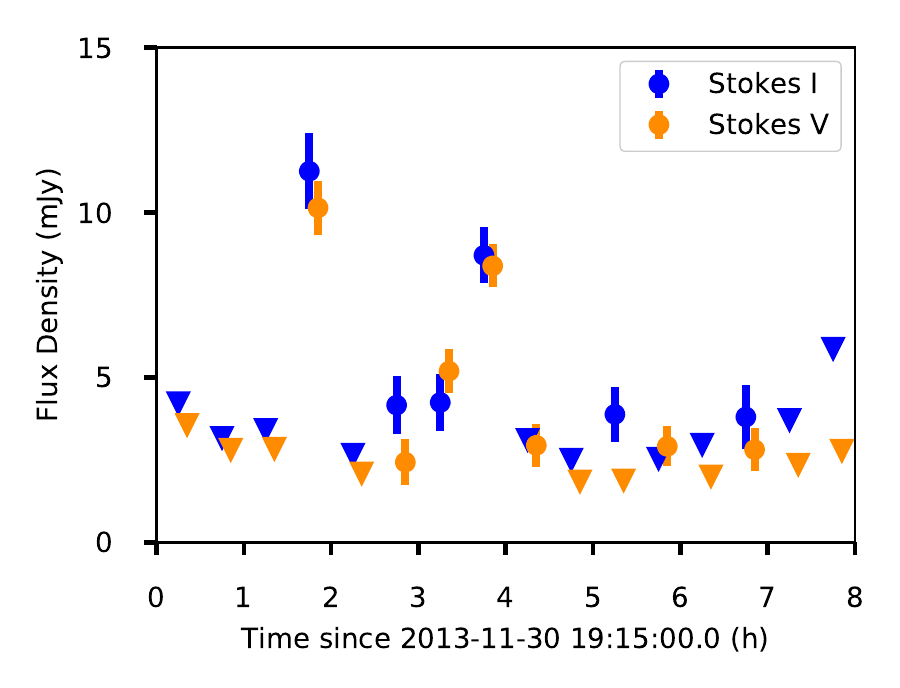}
    \caption{Light curve of KPNO-Tau-14 over the observation with a time resolution of 30 min. Stokes I and Stokes V measurements are shown by the blue and orange data points respectively. The error bars are given by the root-mean-square noise $\sigma_{rms}$ in the image. The Stokes V measurements are offset by +0.1 h with respect to the Stokes I measurements for better readability. Note that the absolute flux calibration error is not taken into account in the error bars as any absolute flux calibration offset should be the same for the whole observation. $3\sigma$ upper limits are shown by the triangles for the times in which no emission was detected from the source. See Appendix D for the flux density measurements and calculated polarization fractions.}
    \label{KPNO-Tau_14_light_curve}
\end{figure}{}

\begin{figure}
    \centering
    \includegraphics[width=\hsize]{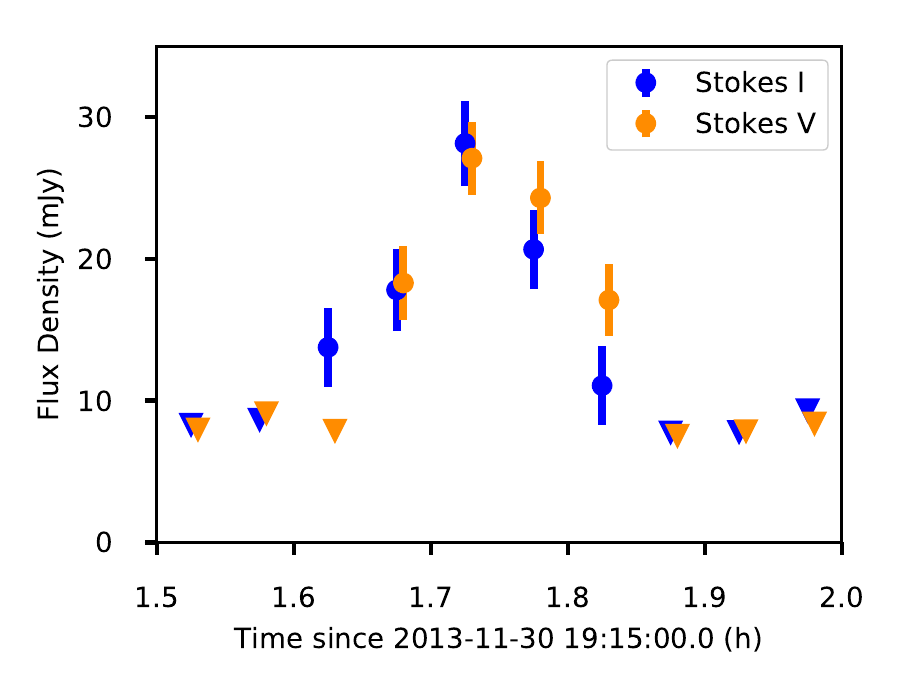}
    \caption{Light curve of the first burst detected from KPNO-Tau-14 with a time resolution of 3 min. Stokes I and Stokes V measurements are shown by the blue and orange data points respectively. The error bars are given by the root-mean-square noise $\sigma_{rms}$ in the image. The Stokes V measurements are offset by +0.005 h with respect to the Stokes I measurements for better readability. Note that the absolute flux calibration error is not taken into account in the error bars as any absolute flux calibration offset should be the same for the whole observation. $3\sigma$ upper limits are shown by the triangles for the times in which no emission was detected from the source. See Appendix D for the flux density measurements and calculated polarization fractions.}
    \label{KPNO-Tau_14_1st_flare_light_curve}
\end{figure}{}

\begin{figure*}
    \includegraphics[width=\textwidth]{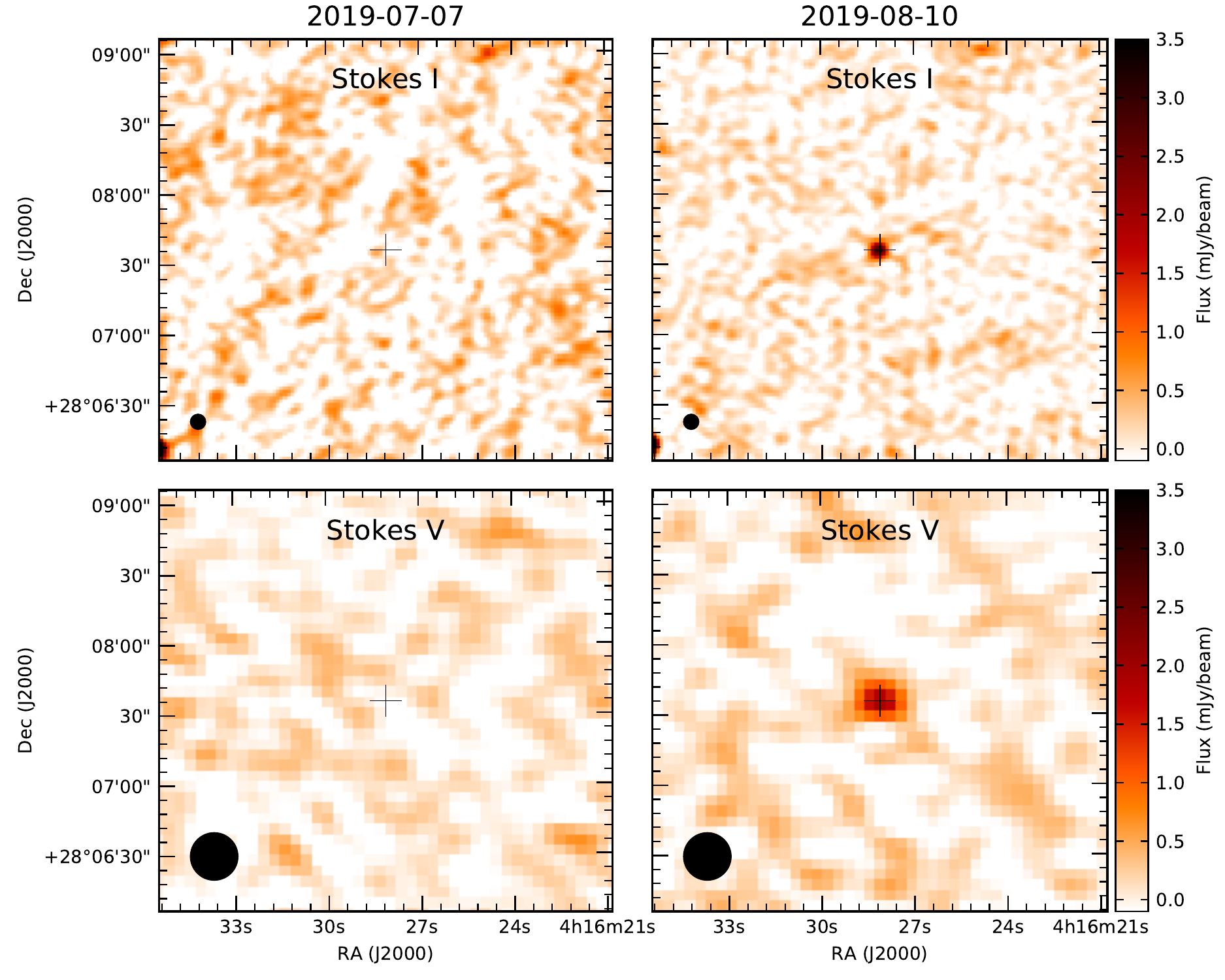}
    \caption{Stokes I (top) and Stokes V (bottom) images of LkCa~4 obtained from LOFAR at two epochs: 2019 July 07 (left) and 2019 Aug 10 (right). The restoring beams used for the images were $6\arcsec \times 6\arcsec$ and $20\arcsec \times 20\arcsec$ for the Stokes I and Stokes V images respectively. The optical position of LkCa 4 obtained from Gaia DR2 \citep{Gaia2018} is indicated by the cross.}
    \label{LkCa_4_images}
\end{figure*}

\section{Results}
\subsection{KPNO-Tau~14}

In the 2013 observations, emission was detected from the position of KPNO-Tau~14 with a peak flux density (Stokes I) of $3.52 \pm 0.73\ \mathrm{mJy\ beam^{-1}}$, as seen in Fig. \ref{KPNO-Tau_14_images}. In addition, emission was also detected with a peak flux density of $3.28 \pm 0.67\ \mathrm{mJy\ beam^{-1}}$ in Stokes V, showing that the emission is highly circularly polarised with a polarisation fraction of $80_{-17}^{+13} \%$. Given that the source is unresolved, we take the peak flux density to be the total flux density. 

The error in the flux density measurement, $ \sigma_{S_{\nu}}$, and subsequent flux density measurements in this paper, was taken to be a combination of the root-mean-square noise $\sigma_{rms}$ of the image and the absolute flux calibration error, found by \citet{Shimwell2019} to be 20\% for LOFAR after applying a correction factor based on other surveys: $ \sigma_{S_{\nu}} = \sqrt{\sigma_{rms}^2 + (0.20 \times S_{\nu})^2}$. The polarization fraction values and errors for this and all subsequent flux density measurements were found as follows. Gaussian distributions were generated for the Stokes I and Stokes V values, with a mean of the measured flux value $S_{\nu}$ and a standard deviation of $ \sigma_{S_{\nu}}$. The values in the Stokes V distribution were divided by the values in the Stokes I distribution to generate a distribution of polarization fraction values. This polarization fraction distribution was then truncated above 100\% to remove any values higher than this. The value of the polarization fraction was then taken to be the median of this truncated distribution and the errors give the 68\% confidence interval.

To investigate time variability, the direction-dependent calibrated data was divided into time bins of 30 minutes. Each period was then imaged using WSClean in Stokes I and V. To check that the flux density scale calibration was accurate for all the images produced, the flux densities of several nearby compact bright sources in the field were measured in each image using PyBDSF and their light curves were plotted to check for any variation in the measured flux densities (see Appendix A for the light curves). Over the 8 hours, the flux densities of these sources varied with an average standard deviation of 12\% relative to the mean flux density of each source. Given that this is within the typical absolute flux calibration error for LOFAR, we concluded that the flux density scale calibration was accurate for all of the images.

It was found that the flux density from KPNO-Tau 14 varied significantly over the 8 hours of observation. This can be seen in the light curve in Fig. \ref{KPNO-Tau_14_light_curve} where 2 bursts can be clearly seen. In particular, the first burst, 1.5 - 2 hours after the start of our observation, is very strong with a flux density of $11.25\pm1.15$ mJy. Note that the errors for this light curve and the subsequent light curves in this paper do not take into account the absolute flux calibration error as any absolute flux calibration offset should be the same for the whole observation.

In order to study the first burst in more detail, the data was further sub-divided into 3-minute time bins and plotted in the light curve seen in Fig. \ref{KPNO-Tau_14_1st_flare_light_curve}, showing that the duration of the first burst was approximately 15 minutes and the flux density peaked at $28.15\pm3.02$ mJy. It should be noted that the true maximum flux density may be higher but due to our time binning, this may have been averaged out. Unfortunately, we cannot obtain a higher time resolution as the signal-to-noise would be too poor.

While the second burst seen in Fig. \ref{KPNO-Tau_14_light_curve} is not as strong as the first, it has a longer duration, lasting from 2.5 - 4.5 hrs after the start of the observation, and peaking at  $8.70\pm0.87$ mJy. Some varying low-level emission is then detected during the rest of the observing window. The emission is highly circularly polarised throughout, with the exception of that detected at 5.0 - 5.5 hrs, and is consistently left-hand circularly polarised (see Appendix B). Both of the major bursts detected are strongly circularly polarised, with polarisation fractions of $88_{-9}^{+8} \%$ and $91_{-8}^{+6} \%$ respectively.

Based on the distance to the source, and assuming the emission is isotropic, i.e.\ not beamed, the first and largest burst had a peak luminosity of at least $(7.8\pm1.0) \times 10^{17}\ \mathrm{erg\ s^{-1}\ Hz^{-1}}$, while the second smaller burst had a peak luminosity of at least $(2.4\pm0.3) \times 10^{17}\ \mathrm{erg\ s^{-1}\ Hz^{-1}}$. To calculate the corresponding brightness temperatures, we assume, as indicative, the size of the emission region to be a circular disk with a radius equal to the radius of the star. From the \citet{Baraffe2015} evolutionary models for TTSs, a $0.1 M_{\sun}$ star with an age of 1~Myr should have a radius of $\sim 1 R_{\sun}$. This would mean that the brightness temperatures of the first and second bursts were at least $(5.8\pm0.7) \times 10^{14}\ \mathrm{K}$ and $(1.8\pm0.2) \times 10^{14}\ \mathrm{K}$ respectively. However, given that the emission region is likely much smaller than the entire stellar surface, we are almost certainly underestimating the actual brightness temperatures.

% Need to do errors for luminosity and brightness temperature). 
% I think you can get away with not doing these since they are very dependent on what size you assuming for the emission region, etc. Tom

Finally, we remark that the presence of strong circular polarisation and the transient nature of the emission implies that the emission is associated with KPNO-Tau~14 and not a background extragalactic source for example. Note also that no emission was detected in either Stokes I or V from KPNO-Tau~14 in the later pointing (P068+26) made in 2019, as seen in Fig. \ref{KPNO-Tau_14_images}, with a $3\sigma$ upper limit on the flux density in Stokes I of $\leq 0.39$ mJy.

\begin{figure}
    \centering
    \includegraphics[width=\hsize]{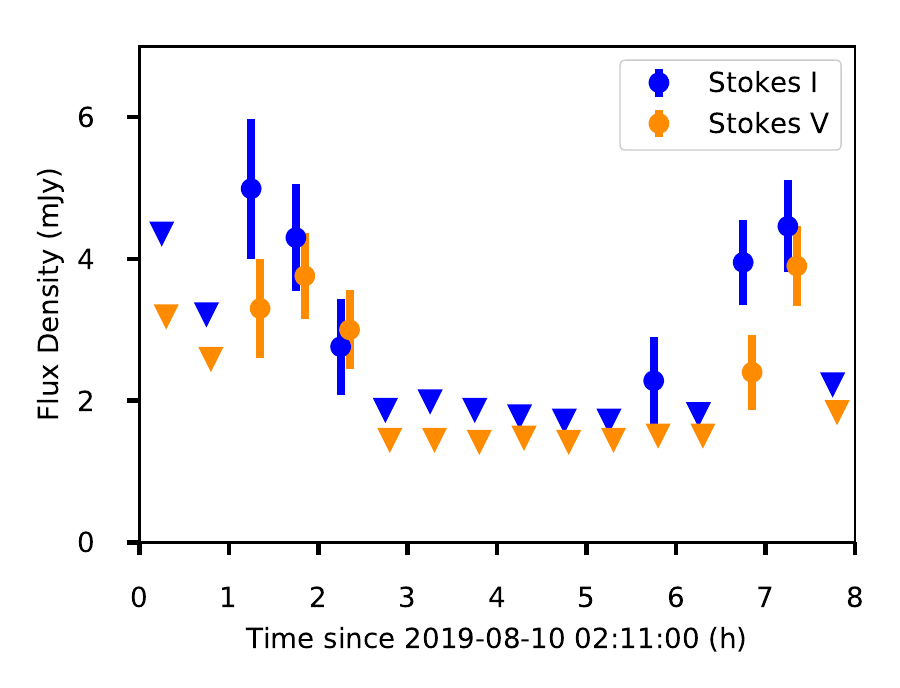}
    \caption{Light curve of LkCa~4 over the observation with a time resolution of 30 min. Stokes I and Stokes V measurements are shown by the blue and orange data points respectively. The error bars are given by the root-mean-square noise $\sigma_{rms}$ in the image. The Stokes V measurements are offset by +0.1 h with respect to the Stokes I measurements for better readability. Note that the absolute flux calibration error is not taken into account in the error bars as any absolute flux calibration offset should be the same for the whole observation. $3\sigma$ upper limits are shown by the triangles for the times in which no emission was detected from the source. See Appendix D for the flux density measurements and calculated polarization fractions.}
    \label{LkCa_4_light_curve}
\end{figure}{}

\subsection{LkCa~4}
During a LoTSS run on August 10 2019, emission was found from LkCa~4 (pointing P064+29) with a peak flux density (Stokes I) of $3.12\pm0.65\ \mathrm{mJy\ beam^{-1}}$, as seen in Fig. \ref{LkCa_4_images}. In Stokes V, the corresponding value was $1.92\pm0.42\ \mathrm{mJy\ beam^{-1}}$, indicating a highly circularly polarised source with a polarisation fraction of $60_{-15}^{+19} \%$. Again, as the source is unresolved, we take the peak flux density to be the total flux density.

As with KPNO-Tau~14, in order to detect changes in the emission, the direction-dependent calibrated data was divided into 30-minute time bins, each of which was then imaged using WSClean in Stokes I and V. The flux densities of several nearby compact bright sources in the field were measured in each image to check for that the flux scale calibration was accurate for all of the images used. Over the 8 hours, the flux densities of these sources varied with an average standard deviation of 8\% relative to the mean flux densities of each source. Since this was well within the typical absolute flux calibration error for LOFAR, we concluded that the flux density scale calibration was accurate for all of the images (see Appendix A for the light curves).

\begin{figure}
    \centering
    \includegraphics[width=\hsize]{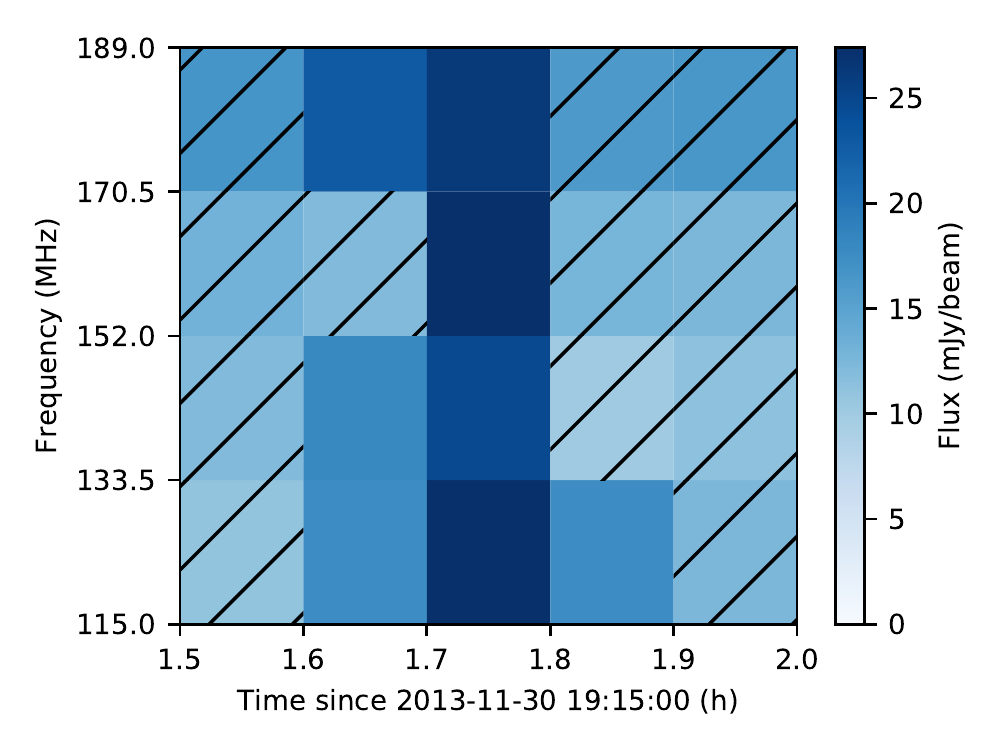}
    \caption{Dynamic spectrum of the first burst detected from KPNO-Tau~14. The burst was divided into time-frequency bins with a time resolution of 6 min and a frequency resolution of 18.5 MHz. The source was then imaged in each bin and its flux density was measured. $3\sigma$ upper limits are shown for the bins in which no emission was detected, indicated by the hatched lines. The mean noise of the images used for this spectrum was 4.4 mJy.}
    \label{freq_analysis}
\end{figure}

As seen from the light curve in Fig. \ref{LkCa_4_light_curve}, there were two bursts observed during the 8 hour run. The first lasted from 1.0 - 2.5 hours after the start of the observing period, peaking at $4.99\pm0.99$ mJy. The second lasted from approximately 6.5 - 7.5 hours, peaking at $4.46\pm0.64$ mJy. Both bursts are circularly polarised in a left-handed sense.

Based on the distance to the star, the peak luminosity of the first and second bursts are $\sim (1.0\pm0.2) \times 10^{17}\ \mathrm{erg\ s^{-1}\ Hz^{-1}}$ and $\sim (0.9\pm0.1) \times 10^{17}\ \mathrm{erg\ s^{-1}\ Hz^{-1}}$ respectively. Using a circular disk with a radius equal to the stellar radius $\sim 2.3\, R_{\sun}$  \citep{Gully-Santiago2017}, as an indicative size of the source, the corresponding brightness temperatures are $\sim (1.6\pm0.3) \times 10^{13}\ \mathrm{K}$ and $\sim (1.4\pm0.2) \times 10^{13}\ \mathrm{K}$ respectively. Similar to KPNO-Tau 14, however, it should be noted that the brightness temperatures are almost certainly much higher as the emission region is likely far smaller than the stellar disk.

We note that no emission was detected from LkCa~4 in the 3 other relevant LoTSS pointings, P061+29, P062+26, and P065+26, for which the corresponding $3\sigma$ Stokes I upper limits are $\leq 0.69$ mJy, $\leq 0.72$ mJy, and $\leq 0.84$ mJy respectively.

To check if the emission was related to the rotation period of the star, the rotation phases observed at each of the observations were compared. The phases were calculated based on the barycentric Julian dates (BJD) of the observations using the following expression
\begin{equation}
    E = \frac{BJD - 2458662.7}{3.37}
\end{equation}
where E is the rotational phase of the star, $3.37 \pm 0.01$\,d is the rotational period of LkCa\,4 \citep{Grankin2008}, and 2458662.7 is the barycentric Julian Date at the start of the first observation (P061+29) which we have arbitrarily chosen to correspond to a rotation phase of 0.0. The calculated rotation phases observed in each of the observations are listed in Table \ref{LkCa_4_observations}.

The two bursts observed in the LoTSS pointing P064+29 occur at phases $\approx 0.72 - 0.74$ and $\approx 0.79 - 0.80$ respectively. Therefore, the rotation phase corresponding to the first burst is observed in the observation of LoTSS pointing P065+26 which covers the phase range $0.67 - 0.77$. Note that there is an uncertainty in the phase of this observation relative to the observation of P064+29 of $\pm0.03$ due to the uncertainty in the rotation period of the star and the period of 1 month that elapsed between the two sets of observations. Given that the source is not seen in this observation it appears that the emission is not related to the rotation phase. Although it is important to note that the two sets of observations are a month apart and the bursting mechanism may be episodic.

In addition, detecting periodicity is also made difficult by potentially variable conditions in the source region. For example, for ECM emission variable maser conditions can lead to substantial variations in flux. In the case of CR Draconis,  \citet{Callingham2021} detected flux density variability of at least an order of magnitude, which was proposed to be due to `hotspots' in the stellar magnetosphere, containing unusually dense and/or hot plasma, creating highly localised and variable emission. CM emission is generally also highly beamed, which may affect the ability to detect it. If the emission beam is frequently aligned away from our line-of-sight it will not be detected, making it difficult to identify any periodicity.

\subsection{Frequency Analysis}\label{Frequency_Analysis}
For the first burst from KPNO-Tau 14, the emission is sufficiently bright that we can construct a dynamic spectrum. This was done by binning the data for the burst according to time and frequency and then imaging the region around KPNO-Tau 14 in Stokes I in each time-frequency bin and measuring the flux density of KPNO-Tau 14 in each image. The resulting dynamic spectrum is seen in Fig. \ref{freq_analysis}, with a time resolution of 6 min and a frequency resolution of 18.5 MHz.

Using this dynamic spectrum, we can analyse the spectral properties of the burst and search for frequency drift. It can be seen that the burst occurs over the entire bandwidth of the observation. The emission appears to be flat over the bandwidth, with a spectral index at 1.75 h of $-0.03 \pm 0.61$, and the flux density seems to rise and fall similarly across the bandwidth with no frequency drift apparent. However, it should be noted that the time and frequency resolution is very low, and so we cannot rule out any finer time-frequency structure which might be present. In addition, due to the low signal-to-noise ratio and non-detections in many of the time-frequency bins, significant changes in the spectrum of the burst could be hidden by the noise.

Unfortunately, for the other radio bursts detected, the flux density is not bright enough to obtain dynamic spectra due to the low signal-to-noise ratio.

\subsection{Searching for Linear Polarisation}
To search for linear polarisation, rotation measure (RM) synthesis \citep{Brentjens2005} was used. This is required in order to correct for any Faraday rotation in the source, and/or from the foreground, an effect where the plane of linear polarisation rotates as it propagates through a magnetised plasma. The amount of rotation is frequency dependent:
\begin{equation}
    \chi (\lambda) = \chi_0 + \phi\lambda^2
\end{equation}
where $\chi (\lambda)$ is the observed final angle, $\chi_0$ is the initial angle and $\phi$ is the Faraday depth, which depends on the average electron density and longitudinal magnetic field 
$ \phi \propto \int n_{\mathrm{e}} \mathbf{B} \cdot d\mathbf{r} $.
As a result, when integrating a signal over a sufficiently wide bandwidth, the signal can become depolarised. This effect is particularly strong for LOFAR, as it operates at such long wavelengths and has a large fractional bandwidth.

RM synthesis can get around this problem by imaging the  Q and U images in narrow frequency channels and then performing a Fourier transform from the linearly polarised emission as a function of $\lambda^2$ to the linearly polarised emission as a function of $\phi$.

When RM synthesis was done at the position of KPNO-Tau-14, no significant linearly polarised emission was found at any Faraday depth in the Faraday dispersion spectrum. The noise level in the Faraday dispersion was $189\ \mathrm{\mu Jy\ beam^{-1}\ RMSF^{-1}}$, where RMSF is the RM spread function, a point spread function in Faraday space due to finite sampling in $\lambda^2$-space. This gives a $3\sigma$ upper limit of $<17\%$ on the linear polarisation fraction.

Similarly for LkCa~4, no evidence for linearly polarised emission was found. The noise level was $237\ \mathrm{\mu Jy\ beam^{-1}\ RMSF^{-1}}$, giving a $3\sigma$ upper limit on the linear polarisation fraction of $<35\%$.

However, this does not rule out the possibility of linear polarization as we may be unable to detect it due to lack of sensitivity. For example, \citet{Callingham2021} detected linear polarization from similar emission in CR Draconis with a linear polarization fraction of $<10\%$, which would be too faint to detect in the emission we observe.

\section{Discussion}

\subsection{What is the Relevant Emission Mechanism?}
Both sources show a high level of circular polarisation in the bursts observed. In addition, their brightness temperatures are at least on the order of $\sim 10^{13} - 10^{15}\ \mathrm{K}$. This indicates a coherent emission mechanism, as incoherent mechanisms are limited to $10^{11} - 10^{12}\ \mathrm{K}$ since above this, the source would be rapidly cooled by inverse Compton scattering \citep{Kellermann1969}. 

Assuming we are dealing with coherent radiation, the options are either plasma emission, as seen for example in the Sun from Type II radio bursts \citep{Hariharan2014}, or ECM emission which is associated with auroral emitters such as planets, brown dwarfs and main sequence M dwarfs 
\citep{Lynch2017a, Pineda2017, Villadsen2019, Zic2019, Vedantham2020,Callingham2021} and possibly Type IV solar radio bursts \citep{Treumann2011}. 

\subsection{Plasma Emission}
Plasma emission occurs at the local plasma frequency, $\nu_p$, and its second harmonic, but rarely if ever at higher harmonics \citep{Dulk1985}:
\begin{equation}
    \nu_p = \sqrt{n_e e^2 / \pi m_e} \approx 9 \sqrt{n_e}\ \mathrm{kHz}
\end{equation} 
where $n_e$ is the local electron number density in $\rm cm^{-3}$. Here we have assumed for simplicity that no magnetic field is present. At the fundamental frequency, plasma emission is expected to be polarised. However, the second harmonic is usually not polarised or only weakly polarised \citep[$<50 \%$;][]{Melrose1978}. Given the high polarisation fraction of the emission in both our sources, second harmonic plasma emission can therefore be ruled out. 

The origin of plasma emission involves the production of longitudinal waves, known as Langmuir waves, in a plasma. These are usually generated by the injection of a hot plasma into a cooler ambient plasma. Some of the energy in the Langmuir waves is subsequently converted into electromagnetic waves.

Plasma emission, especially at higher frequencies, can suffer from free-free absorption due to overlying lower-density plasma blanketing the plasma in the source region \citep{Dulk1985}. However, at low frequencies this is generally not an issue as the free-free optical depth is proportional to frequency: $\tau_{\mathrm{ff}} \propto \nu^2$

The maximum possible brightness temperature that can be generated by plasma emission can be estimated from the effective temperature of the Langmuir waves $T_{\mathrm{L}}$, defined by \citep{Dulk1985}:
\begin{equation}
    W_{\mathrm{L}} = \int \frac{k_{\mathrm{B}}T_{\mathrm{L}}}{(2\pi)^3} d^3 \mathbf{k}
\end{equation}
where $W_{\mathrm{L}}$ is the energy density of the Langmuir waves. It has been shown that the saturated value of $W_{\mathrm{L}}$ is $10^{-5}$ times the energy density of the background plasma i.e. $W_{\mathrm{L}} \approx 10^{-5} k_{\mathrm{B}} T$ \citep{Benz1993}, where $T$ is the temperature of the background plasma. From this, the limiting value of $T_{\mathrm{L}}$ can be shown to be \citep{Dulk1985}:
\begin{equation}
    T_{\mathrm{L}} \lesssim 10^8 \frac{\rm {v_o^2}}{c^2} \frac{\rm{v_o}}{\omega_p} T
\end{equation}
where ${\rm v_o}$ is the velocity of the fast electrons in the hot plasma component and $\omega_p = 2\pi \nu_p$. Given that the brightness temperature of the plasma emission is limited to the effective temperature of the Langmuir waves, this also gives an upper limit on the brightness temperature.

From X-ray observations \citep{Gudel2007}, the average electron temperature in the corona of KPNO-Tau 14 is known to be $1.65 \times 10^{7}\ \mathrm{K}$. Using this as an estimate for $T$ and assuming a value of $v_o$ similar to those seen in the Sun, $\approx 0.1 c - 0.5 c$, the brightness temperature is therefore limited to $T_{\mathrm{b}} \lesssim 10^{15}\ \mathrm{K}$. While this is just about consistent with the estimated brightness temperatures of the bursts detected, it should be noted that  those values were estimated based on an emission region equal to the size of the stellar disk and so are only lower limits. In reality, the emission region is likely much smaller and therefore the brightness temperature much higher.

This suggests that plasma emission is unlikely to be able to explain the bursts observed. Additionally, if the emission was coming from the entire stellar disk, this would imply the emission was due to flaring loops occurring across the entire stellar disk. If this were the case, we would not expect to obtain such high polarization fractions as flaring loops from regions with oppositely directed magnetic fields would generate emission of oppositely handed polarization which would cancel each other out. Thus a high polarization fraction such as that observed would require a smaller emission region.

Given that the temperature of the corona used is only an average, it is possible that the temperature could be higher in smaller regions of the corona. However, this runs into the issue again that the brightness temperature would then be required to be much higher in a smaller emission region and this would in turn require an even higher plasma temperature.

For LkCa~4, unfortunately no estimates of the ambient coronal temperature are available. However, assuming it is within the range typically seen for T Tauri stars, i.e.\ $10^6 - 10^7\ \mathrm{K}$ \citep{Gudel2007}, and taking the same estimates for $v_o$ as before, the brightness temperature for plasma emission is limited to $T_{\mathrm{b}} \lesssim 10^{14} - 10^{15} \ \mathrm{K}$. Again, this is marginally consistent with the estimated brightness temperatures of the bursts. Similar to KPNO-Tau 14 however, these estimated values are only lower limits based on an emission region equal in size to the stellar disk. As the real emission region is likely much smaller and therefore the brightness temperature likely much higher, it seems unlikely that plasma emission can explain the bursts.

\begin{figure}
    \centering
    \includegraphics[width=0.5\textwidth]{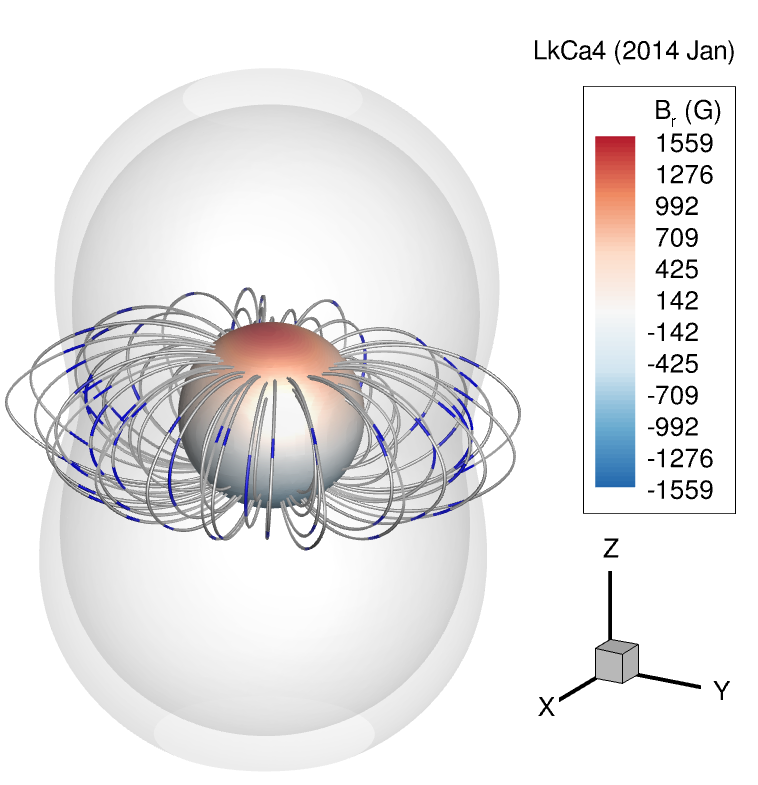}
    \caption{3D magnetic map of LkCa 4, obtained by extrapolating the surface magnetic field maps obtained by \citet{Donati2014}. The positive magnetic field strength values indicate where the magnetic field is pointing outwards from the surface while the negative values indicate where it is pointing inwards towards the surface. The model used to calculate this magnetic map assumes that the field lines become fully radial at $3.4\ \mathrm{R_*}$. The grey surfaces indicate the region in the magnetosphere where the magnetic field strength is $40 - 70\ \mathrm{G}$. The locations along the magnetic field lines where the magnetic field strength is in this range are indicated by the blue line segments. Note that only the closed magnetic field lines are shown in this figure.}
    \label{LkCa_4_magnetic_map}
\end{figure}

\subsection{ECM Emission}
ECM emission is produced when radiation is amplified by stimulated emission at the local cyclotron frequency and its harmonics \citep{Dulk1985}:
\begin{equation}
    \nu_c = 2.8 B\ \mathrm{MHz}
\end{equation}
where $B$ is the local magnetic field strength in G. In order for ECM emission to occur, the following conditions must be met: (a) a population inversion must exist in the electron energy distribution so that amplification by stimulated emission can occur, (b) the plasma frequency must be less than the cyclotron frequency i.e. either a strong magnetic field or a low-density plasma is required. Note that while the bandwidth of an individual pulse of ECM radiation is small, the frequency range of the emission as a whole can be quite large depending on the size of the emission regions and the range of magnetic field strengths that it spans.

One issue for ECM emission is the absorption of the emission by overlying plasma due to gryoresonance absorption \citep{Dulk1985}. This is particularly an issue for fundamental ECM emission as this is expected to be strongly absorbed at the second gyroresonance layer (where $B = B_{\mathrm{src}}/2$). One way to reduce this problem is to assume the emission is produced at the second harmonic. In this case, the emission only has to cross the third harmonic layer (where $B = 2B_{\mathrm{src}}/3$). Alternatively, this issue can be avoided if the emission is generated in a low density cavity within the magnetosphere, as this would give rise to a lower optical depth \citep{Villadsen2019,Zic2019}.

If we are dealing with ECM emission, then we can calculate the local magnetic field strength in the source region. Given the observed LOFAR frequency range, and assuming the that it is fundamental ECM emission, the emission must be from a region with a field strength ranging from 40\,G to 70\,G.

For LkCa 4, \citet{Donati2014} showed that the magnetosphere consists primarily of a  $1.6\ \mathrm{kG}$ magnetic dipole component. Assuming a dipolar magnetosphere as an approximation, the distance of the emitting region from the star can then be estimated. This gives an emission region location at a height of $\approx$ 2\,${\rm R_*}$. A more accurate estimate can be obtained by extrapolating the surface magnetic maps obtained by \citet{Donati2014} to obtain a 3-D map of the magnetosphere of LkCa~4, shown in Fig. \ref{LkCa_4_magnetic_map}. The grey surfaces indicate the region in the magnetosphere where the magnetic field strength is $40 - 70\ \mathrm{G}$. The locations along the closed magnetic field lines where the magnetic field strength is in this range are indicated by the blue line segments, suggesting that the emission is from a region at a height of $\approx 2\  \mathrm{R_*}$.

No surface field strength estimates are available for KPNO-Tau 14. Assuming however it is of the same order of magnitude as LkCa 4, then the emission should be from similar height.

However, we are limited by the observation bandwidth. The frequency range of the ECM emission could extend well beyond that of LOFAR, which would imply larger emission regions and therefore a larger range of magnetic field strengths covered. For example, in the case of Earth and Jupiter, ECM emission has been observed with a frequency range of an order of magnitude or more \citep{Melrose2017}.

Finally, it is worth repeating that for ECM emission to occur, the condition $\nu_p \lesssim \nu_c$ must be met \citep{Dulk1985}. Therefore an upper limit on the plasma density for both stars in their emitting regions can be obtained: $n_e \lesssim 10^8\ \mathrm{cm^{-3}}$.

\subsubsection{Could the ECM Emission Be Generated by Coronal Loops?}
One method of producing ECM emission is through a flaring coronal loop in the magnetosphere of the star. Heated thermal plasma is injected into the loop and creates a loss-cone distribution through magnetic mirroring at either end \citep{Dulk1985}. 

The brightness temperature generated by such a loop would depend on its size. To determine the dimensions of the loop needed to attain the observed brightness temperatures, we use the model given by \citet{Vedantham2020}, which assumes a continuously operating loss-cone maser. This model suggests coronal loops of $10^3\ \mathrm{R_{*}}$ in the case of KPNO-Tau 14, and $10^2\ \mathrm{R_{*}}$ for LkCa~4. These values are far too large to be realistic suggesting the emission cannot be generated by this mechanism.

\subsubsection{Co-rotation Breakdown}
ECM emission can be produced via the breakdown of co-rotation between the plasma surrounding the star and its magnetic field. This results in field-aligned currents which accelerate electrons into the corona and chromosphere of the star, creating the population inversion needed for ECM emission. Effectively this is a scaled-up version of the mechanism driving auroral radio emission in the case of Jupiter \citep{Zarka1998} and ultracool dwarfs \citep{Nichols2012}. More recently, this mechanism has been proposed to explain the LOFAR detected main-sequence M dwarf CR~Draconis \citep{Callingham2021}. 

Both KPNO-Tau 14 and LkCa 4 are fast rotators, with rotation periods of 1.86 days \citep{Scholz2018} and 3.37 days \citep{Grankin2008} respectively. In addition, both show evidence of intense stellar activity, given their X-ray fluxes and, in the case of LkCa~4, high starspot coverage. Certainly such activity could provide a source of plasma to drive the emission. In addition, weak-line T Tauri stars are expected to have intense ionized winds with mass-loss rates, on the order of $10^{-12} - 10^{-9}\ \mathrm{M_{\odot}}$ \citep{Vidotto2017}, i.e.\ many orders of magnitude stronger than the Solar Wind. Again such winds may prove to be a source of the plasma.

In the case of ultra-cool dwarfs and Jupiter, auroral radio emission is thought to occur at a wide range of heights along magnetic field lines from near the surface to several radii of the object \citep{Nichols2012,Zarka1998}, giving rise to a wide frequency range of emission. If this is also the case for KPNO-Tau 14 and LkCa 4, this would be consistent with the inferred emission region at a height of $\approx 2 R_*$ for LkCa 4. One would then also expect to see ECM emission at a wide range of frequencies, above and below that observed with LOFAR. For LkCa 4, with a measured polar magnetic field strength of 2 kG \citep{Donati2014}, this would imply ECM emission could, in principle, be detected at frequencies of up to 5.6 GHz but see Section \ref{ECMatHigherFrequencies}.

\subsubsection{Star-planet interaction}
Another possible method for generating ECM emission is through the interaction of a stellar wind with the magnetosphere of an orbiting exoplanet. This is analogous to what happens in the Solar System with the giant planets \citep{Farrell1999}. For example \citet{Vidotto2010} and \citet{Lynch2018} predicted that hot Jupiters orbiting young stars should produce the highest flux density of radio emission of any type of exoplanet.

From magnetohydrodynamical simulations of the stellar wind of the host star, \citet{Vidotto2017} predicted that the exoplanet V830 Tau b, a 2 Myr-old hot Jupiter orbiting a weak-line T Tauri star at 0.057 au, should produce an average radio flux density of 6 to 24 mJy, for a planet with a radius of 1 - 2 Jupiter radii. However, the flux density would not be constant, as it would vary over the planet's orbit as it moves through different regions of the stellar wind. This radio emission would be emitted at frequencies between 28 - 240 MHz, depending on the planetary magnetic field strength (10 - 100 G), meaning that such emission could potentially be detectable with LOFAR.

Given that V830 Tau is a weak-line T Tauri star, like KPNO-Tau 14 and LkCa 4, and is at a similar distance (147 pc), a hot Jupiter orbiting these sources could be able to generate similar levels of radio emission. In order for such emission to be observed using LOFAR at HBA frequencies, the surface magnetic field strength of the planet would need to be $\gtrsim 50\ \mathrm{G}$.

Alternatively, another method of generating radio emission through star-planet interaction is through a sub-Alfv\'enic interaction, similar to that seen in the Jupiter-Io interaction which causes the Io-dependent decametric (Io-DAM) radio emission, in which the stellar magnetosphere couples with an orbiting body \citep{Saur2013, Turnpenney2018}. Magnetic energy flux is carried by Alfv\'en waves from the orbiting exoplanet and back towards the star, generating radio emission along the field line connecting the star and the planet. This was recently proposed to cause the radio emission seen with LOFAR in the nearby M dwarf GJ 1151 \citep{Vedantham2020}, and later confirmed through detection of a compatible exoplanet using the radial velocity method \citep{Mahadevan2021}.

To check if this is a plausible explanation for the emission that we have detected, we calculated the theoretical estimate for the power of the Alfv\'en waves generated by the interaction based on the following expression \citep{Zarka2007,Saur2013}:
\begin{equation}
    P_{\mathrm{th}} = \frac{1}{2} R^2 v_{\mathrm{rel}}(B \sin \theta)^2 M_A
\end{equation}
where $R$ is the radius of the obstacle to the plasma flow (i. e. the planetary magnetosphere), $v_{\mathrm{rel}}$ is the relative velocity of the stellar wind with respect to the planet, $B$ is the stellar magnetic field strength at the location of the planet, $\theta$ is the angle between the stellar magnetic field and $v_{\mathrm{rel}}$, and $M_A$ is the Alfv\'en Mach number at the position of the planet. Details of the model of the stellar wind and magnetic field used are given in Appendix C.

This was then compared with the observationally implied power based on the flux density observed:
\begin{equation}
    P_{\mathrm{obs}} = F \Omega d^2 \Delta \nu \epsilon
\end{equation}
where F is the observed radio flux density, $\Omega$ is the solid angle into which the emission is beamed, $d$ is the distance to the source, $\Delta \nu$ is the total bandwidth of the emission, and $\epsilon$ is the efficiency of conversion of the energy of the v\'en waves to radio emission. The solid angle is estimated to be $\Omega = 1.6\ \mathrm{sr}$, similar to that of the Jovian radio emission \citep{Zarka2004}. The total bandwidth is taken to be the peak cyclotron frequency of the star $\Delta \nu \approx 2.8 B_0\ \mathrm{MHz}$, where $B_0$ is the surface magnetic field strength of the star in G. 

For LkCa~4, again taking a surface magnetic field strength of 1.6 kG, this gives $P_{\mathrm{obs}} \sim 6 \times 10^{27}\ \mathrm{erg\ s^{-1}}$ for the brightest burst observed. No magnetic field strength measurements are available for KPNO-Tau~14. Therefore, we take an estimate of 1 kG which gives $P_{\mathrm{obs}}\sim (2.8\pm0.4) \times 10^{28}\ \mathrm{erg\ s^{-1}}$ for the brightest burst observed. The efficiency of conversion was taken to be $\epsilon = 0.01$, similar to the value seen for the Io-DAM emission.

If, for example, there were a planet orbiting at $10\, \mathrm{R_*}$, it can be shown by using equations (7) and (8), that such values are appropriate to a planet with a magnetosphere of radius $0.6\, \mathrm{R_J}$, where $\mathrm{R_J}$ is the radius of Jupiter, for LkCa~4,  or a planet  with a magnetosphere of radius $2\, \mathrm{R_J}$ for KPNO-Tau 14. Such planets would likely be hot Jupiters. However, due to the many unknown parameters in this calculation, e.g. the solid angle of the emission, the efficiency of conversion, the orbital semi-major axis of the planet, these are only very rough estimates.

Based on these calculations, it seems that a star-planet interaction is at least a plausible explanation for the radio emission observed, either due to a stellar wind interaction or a sub-Alfv\'enic interaction. In order to determine whether or not the emission is of planetary origin, the YSOs would have to be monitored for a longer time to see if there is any periodicity in the radio emission unrelated to the rotation period of the star. A detection of an exoplanet around either source orbiting close to the star would also be a strong piece of evidence in favour of this. Unfortunately, detecting exoplanets in orbit around TTSs is difficult. Due to their high levels of activity, they generate large fluctuations in rotational velocity (RV). This makes it difficult to detect any orbiting exoplanets, even hot Jupiters.

Neither of the two sources are known to host any exoplanets. In the case of LkCa~4, \citet{Donati2014} attempted to detect a hot Jupiter by searching for periodic signals in its RV curve, after filtering out the RV variations due to the activity of the star. However, they were unable to find any signal with a semi-amplitude greater than $0.1\ \mathrm{km\ s^{-1}}$, and concluded that there was no evidence for a hot Jupiter around LkCa~4, although they suggest that it is possible that there could be a planet with a similar orbital period to the rotation period of the star as its RV signal would then be hidden.

\section{Could the ECM Mechanism Contribute Significantly to Higher Frequency Radio Emission from YSOs?} 
\label{ECMatHigherFrequencies}

As stated earlier, until the availability of facilities such as LOFAR and the GMRT, most radio observations of YSOs have been made at cm wavelengths \citep{Anglada2018}. In the latter cases, what is observed on source tends to fall into two categories: either relatively steady (over at least several months) thermal bremsstrahlung emission arising from the ionised component of an outflow \citep{2021NewAR..9301615R} or non-thermal flare-like emission, sometimes with a quasi-continuous background, that can vary over periods even less than an hour \citep[]{Forbrich2017}. The source of this emission is usually attributed to gyro-synchrotron radiation, produced by mildly relativistic electrons with energies around a few MeV. While outflow/wind type thermal emission has only been seen from classical T Tauri stars (or their embedded counterparts) in the solar-mass regime, non-thermal emission is mainly observed from weak-line T Tauri systems, although it has also been seen in less evolved systems \citep{Dzib2015}. In all of these stars high energy processes are believed to be operating, however, in the case of less evolved systems, the non-thermal emission is expected to be absorbed by an overlying plasma \citep{1996ASPC...93..273A}.

As ECM emission has been detected here at metre wavelengths from two YSOs, it is reasonable to ask whether some, or all, of the non-thermal radio flux in the cm band seen in other YSOs could be attributed to the same mechanism? The first point to note is that, if ECM emission is present, we expect to find a very high level of circular polarisation ($\approx\, 100\%$). In addition we would also anticipate very high surface brightness temperature measurements. Only a few detections of circular polarisation from YSOs at cm wavelengths exist and certainly some \citep{1993ApJ...408..660S,1998ApJ...494L.215F} appear consistent with optically thin gyro-synchrotron emission \citep{Dulk1985} where the circular polarisation fraction can reach at most a few tens of a per cent. That this is the appropriate mechanism is then backed up by the expected surface brightness temperature measurements. We are aware of just one YSO, T~Tau~Sb, where approximately 100\% circular polarisation has been observed on source at cm wavelengths \citep{2003A&A...406..957S} and another, EM*~SR~20, in the decimetre regime \citep{2021MNRAS.502.5438P}. In at least the former case, the measured size of the emitting region, in combination with the flux, also suggests a coherent mechanism must be responsible. Further observations are clearly required to ascertain whether the type of emission observed here is common amongst YSOs at low frequencies or whether it is the rarity seen at higher frequencies.

\section{Conclusions}
We have detected flaring radio emission from two weak-line T Tauri stars, KPNO-Tau 14 and LkCa 4, located in the Taurus Molecular Cloud, at $\sim 150\ \mathrm{MHz}$ using LOFAR. The emission detected from both sources is strongly circularly polarised and highly variable over the course of each observation.

Based on the characteristics of the emission, in particular the high brightness temperature ($\gtrsim 10^{13} - 10^{15}\ \mathrm{K}$) and polarisation fraction ($60 - 90 \%$), we conclude that the emission must be due to a coherent emission mechanism, either plasma or ECM emission.

Of the two coherent emission mechanisms, plasma emission seems unlikely as it cannot generate the high brightness temperatures and polarisation fractions observed and so we conclude that ECM must be responsible.

In order to generate ECM emission, a breakdown of co-rotation between plasma surrounding the star and the stellar magnetosphere is the most likely explanation. Given that weak-line T Tauri stars are expected to have high mass-loss rates, this would provide a source of plasma to drive the emission. Such emission would be a scaled up version of the auroral radio emission seen in Jupiter and ultra-cool dwarfs.

Alternatively, ECM emission could also be generated through a star-planet interaction. This could either be an interaction of the stellar wind with the planetary magnetosphere of an orbiting exoplanet, as seen in our own solar system with the giant planets, or a sub-Alfv\'enic interaction of the stellar magnetosphere with the exoplanet, similar to the Jupiter-Io interaction and seen recently in the M dwarf GJ 1151. If a star-planet interaction is responsible, it could provide important information about exoplanetary magnetospheres. Additionally, given the difficulty in detecting exoplanets around YSOs using traditional methods, this could provide an alternative way of detecting young exoplanets.

These detections suggest that low-frequency radio observations are a valuable method for studying YSOs. As mentioned earlier, the Taurus and Perseus molecular clouds have recently been surveyed using LOFAR. Therefore, we may potentially detect similar emission from other YSOs in these clouds as part of this survey and other observations at low frequencies, giving us important information about their magnetic activity and potentially indicating the presence of exoplanets.

\begin{acknowledgements}
AFJ, SJDP and TPR acknowledge funding from the European Research Council (ERC) under Advanced Grant No.\ 743029. AAV acknowledges funding from the European Research Council (ERC) under the European Union's Horizon 2020 research and innovation programme (grant agreement No 817540, ASTROFLOW). This paper is based on data obtained with the International LOFAR Telescope (ILT) under project codes LC1\_001 and LC12\_015. LOFAR \citep{VanHaarlem2013} is the Low Frequency Array designed and constructed by ASTRON. It has observing, data processing, and data storage facilities in several countries, that are owned by various parties (each with their own funding sources), and that are collectively operated by the ILT foundation under a joint scientific policy. The ILT resources have benefitted from the following recent major funding sources: CNRS-INSU, Observatoire de Paris and Université d'Orléans, France; BMBF, MIWF-NRW, MPG, Germany; Science Foundation Ireland (SFI), Department of Business, Enterprise and Innovation (DBEI), Ireland; NWO, The Netherlands; The Science and Technology Facilities Council, UK. This research made use of the Dutch national e-infrastructure with support of the SURF Cooperative (e-infra 180169) and the LOFAR e-infra group. The Jülich LOFAR Long Term Archive and the German LOFAR network are both coordinated and operated by the Jülich Supercomputing Centre (JSC), and computing resources on the supercomputer JUWELS at JSC were provided by the Gauss Centre for Supercomputing e.V. (grant CHTB00) through the John von Neumann Institute for Computing (NIC). This research made use of the University of Hertfordshire high-performance computing facility and the LOFAR-UK computing facility located at the University of Hertfordshire and supported by STFC [ST/P000096/1], and of the Italian LOFAR IT computing infrastructure supported and operated by INAF, and by the Physics Department of Turin university (under an agreement with Consorzio Interuniversitario per la Fisica Spaziale) at the C3S Supercomputing Centre, Italy. This research made use of Matplotlib \citep{Hunter2007}, Scipy \citep{2020SciPy-NMeth}, and of APLpy, an open-source plotting package for Python \citep{APLpy2012}.
\end{acknowledgements}


\bibliographystyle{aa}
\bibliography{Radio_Flares}

%
%-------------------------------------------------------------
%               Appendices have to be placed at the end, after
%                                        \end{thebibliography}
%-------------------------------------------------------------
%\end{thebibliography}

\begin{appendix} %First appendix
\section{Light curves of background field sources}
\begin{figure}[h]
    \centering
    \includegraphics[width=0.5\textwidth]{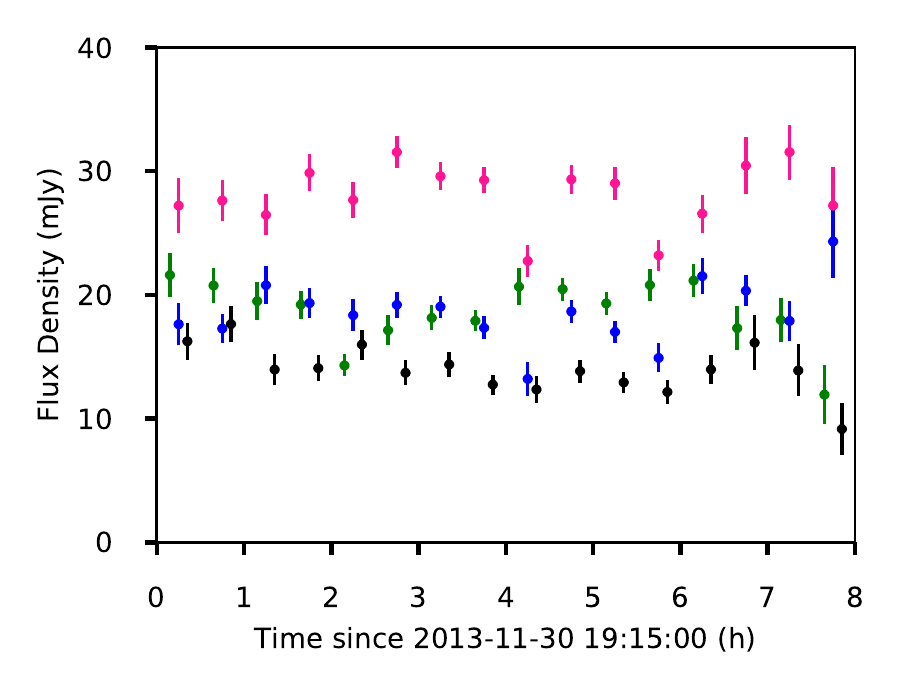}
    \caption{Light curves of the sources in the field of KPNO-Tau 14: J043314+261818 (pink), J043316+261813 (blue), J043349+260753 (green), and J043331+260704 (black). Note that the fluxes of J043349+260753 and J043331+260704 are offset by -0.1 h and +0.1 h respectively for readability. The error bars are given by the fitting error $\sigma_{\mathrm{fit}}$ given by PyBDSF and the root-mean-square noise $\sigma_{\mathrm{rms}}$ added in quadrature: $\sigma = \sqrt{\sigma_{\mathrm{fit}}^2 + \sigma_{\mathrm{rms}}^2}$}
    \label{KPNO-Tau_14_field_source_light_curves}
\end{figure}
\begin{figure}[h]
    \centering
    \includegraphics[width=0.5\textwidth]{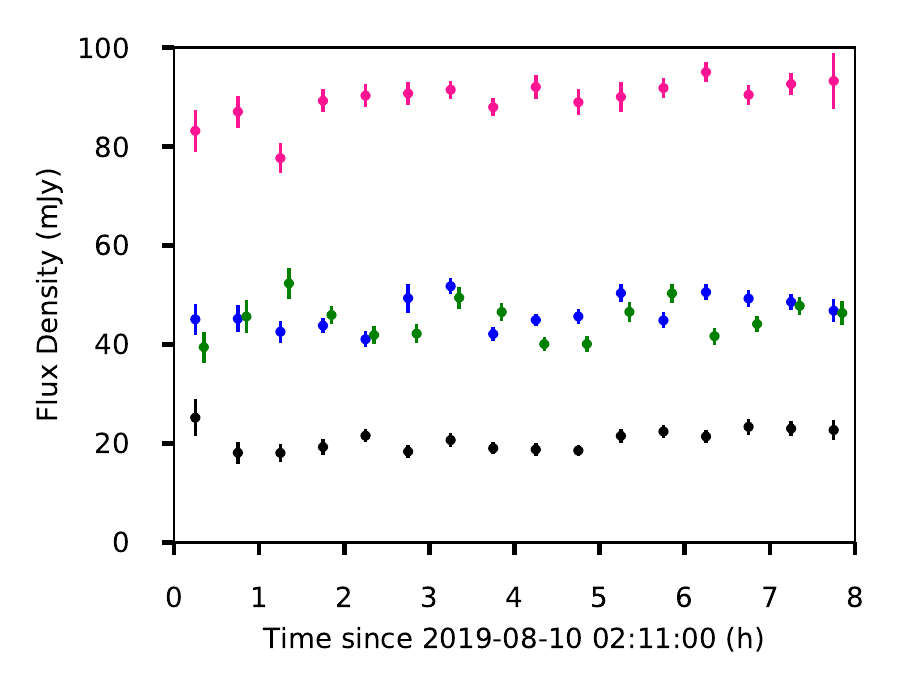}
    \caption{Light curves of the sources in the field of LkCa 4: J041556+281645 (pink), J041732+281048 (blue), J041640+280521 (green), and J041607+281833 (black). Note that the fluxes of J041640+280521 are offset by +0.1 h for readability. The error bars are given by the fitting error $\sigma_{\mathrm{fit}}$ given by PyBDSF and the root-mean-square noise $\sigma_{\mathrm{rms}}$ added in quadrature: $\sigma = \sqrt{\sigma_{\mathrm{fit}}^2 + \sigma_{\mathrm{rms}}^2}$}
    \label{LkCa_4_field_source_light_curves}
\end{figure}

\section{Sign of Stokes V}
There are multiple conventions for defining the Stokes parameters within radio astronomy \citep{Robishaw2018}. The one used by CASA and WSClean is $V = (RR - LL)/2 $. This is different to the official IAU definition which does not contain the factor of 2: $V = RR - LL$. Within pulsar astronomy and optical astronomy, the adopted convention is $V = LL - RR$. There are also other factors that can affect the sign of Stokes V, such as the location of the feed in the antenna (primary, secondary, tertiary focus). Therefore, it is necessary to check by comparing to sources with known circular polarisations e.g. pulsars. This was done by Callingham et al. (in prep.) for LoTSS data, who found that the LoTSS data followed the pulsar convention i.e. $LL - RR$.  Therefore, we can determine that the KPNO-Tau 14 and LkCa 4 emission is left-handed.

\section{Stellar Wind and Magnetic Field Model}
When calculating the theoretical power of the for Alfv\'en waves generated by the sub-Alfv\'enic interaction of the stellar magnetosphere with an orbiting exoplanet, the stellar wind is assumed to have the form of a Parker spiral \citep{Parker1958}, which is given by the solution to the equation:
\begin{equation}
    \frac{u_r(r)^2}{c_s^2} - \ln \left( \frac{u_r(r)^2}{c_s^2} \right) = 4\ln \left(\frac{r}{r_c} \right) + 4\left(\frac{r_c}{r} \right) -3
\end{equation}
where $u_r(r)$ is the radial stellar wind velocity at the radial distance $r$ from the surface of the star. The equation is parameterised by the sound speed:
\begin{equation}
    c_s = \sqrt{\frac{ k_\mathrm{B} T }{m_\mathrm{p}}}
\end{equation}
where $k_\mathrm{B}$ is the Boltzmann constant, $T$ is the temperature of the stellar wind, and $m_p$ is the proton mass. The wind passes through $c_s$ at the critical distance $r_c$:
\begin{equation}
    r_c = \frac{m_\mathrm{p} G M_*}{ 4 k_\mathrm{B} T }
\end{equation}
where $G$ is the gravitational constant and $M_*$ is the mass of the star.

The magnetic field components are given by:
\begin{equation}
    B_r = B_0 \left( \frac{r_0}{r} \right)^2
\end{equation}
\begin{equation}
    B_{\varphi} = B_r \frac{u_\varphi - \Omega r}{v_{\mathrm{rel}}}
\end{equation}
where $\Omega_*$ is the stellar rotational velocity and $u_\varphi$ is the azimuthal stellar wind velocity given by $u_\varphi(r) = (\Omega_* R_*^2)/r$.

\clearpage
\onecolumn
\section{Flux Density measurements}

\begin{table*}[h]
    \caption{Flux densities of KPNO-Tau 14 in Stokes I and V that were used in the light curve in Fig. \ref{KPNO-Tau_14_light_curve}.}
    \centering
    \begin{tabular}{c c c c c}
    \hline 
    Time & Stokes I & Stokes V & Pol. Fraction & Beam \\ 
    (hrs) & (mJy) & (mJy) & (\%) &  \\
    \hline 
    0 - 0.5 & $\leq 4.26$ & $\leq 3.60$ & - & $20\farcs3 \times 3\farcs2$, $-48\degr$ \\ 
    0.5 - 1.0 & $\leq 3.21$ & $\leq 2.85$ & - & $18\farcs7 \times 3\farcs1$, $-52\degr$ \\ 
    1.0 - 1.5 & $\leq 3.44$ & $\leq 2.88$ & - & $16\farcs6 \times 3\farcs1$, $-57\degr$ \\ 
    1.5  - 2.0 & $11.25 \pm 1.15$ & $10.14 \pm 0.82$ & $88_{-9}^{+8}$ & $15\farcs0 \times 3\farcs2$, $-62\degr$ \\ 
    2.0 - 2.5 & $\leq 2.70$ & $\leq 2.13$ & - & $15\farcs0 \times 2\farcs9$, $-68\degr$ \\ 
    2.5 - 3.0 & $4.16 \pm 0.87$ & $2.43 \pm 0.69$ & $57_{-18}^{+20}$ & $13\farcs8 \times 3\farcs1$, $-73\degr$ \\ 
    3.0 - 3.5 & $4.24 \pm 0.87$ & $5.19 \pm 0.67$ & $91_{-10}^{+7}$ & $13\farcs0 \times 3\farcs4$, $-82\degr$ \\ 
    3.5 - 4.0 & $8.70 \pm 0.85$ & $8.38 \pm 0.65$ & $91_{-8}^{+6}$ & $14\farcs2 \times 4\farcs1$, $83\degr$ \\ 
    4.0 - 4.5 & $\leq 3.15$ & $2.94 \pm 0.65$ & $\geq 93$ & $20\farcs4 \times 4\farcs2$, $66\degr$ \\ 
    4.5 - 5.0 & $\leq 2.55$ & $\leq 1.89$ & - & $13\farcs4 \times 4\farcs2$, $62\degr$ \\ 
    5.0 - 5.5 & $\leq 2.55$ & $\leq 1.89$ & - & $7\farcs4 \times 4\farcs2$, $61\degr$ \\ 
    5.5 - 6.0 & $3.88 \pm 0.83$ & $\leq 1.92$ & $\leq 49$ & $6\farcs2 \times 4\farcs2$, $60\degr$ \\ 
    6.0 - 6.5 & $\leq 2.58$ & $2.91 \pm 0.61$ & $\geq 112$ & $5\farcs7 \times 4\farcs3$, $60\degr$ \\ 
    6.5 - 7.0 & $3.80 \pm 0.98$ & $2.81 \pm 0.66$ & $69_{-20}^{+18}$ & $5\farcs5 \times 4\farcs3$, $61\degr$ \\ 
    7.0 - 7.5 & $\leq 3.75$ & $\leq 2.40$ & - & $5\farcs5 \times 4\farcs3$, $62\degr$ \\ 
    7.5 - 8.0 & $\leq 5.91$ & $\leq 2.82$ & - & $5\farcs8 \times 4\farcs3$, $62\degr$ \\ 
    \hline 
    \end{tabular} 
    \tablefoot{The time in the first column refers to the time in hours after the start of the observation (2013-11-30 19:15:00). The polarisation fraction is shown for the times where both Stokes I and V are detected. When only Stokes I or V are detected, $3\sigma$ upper or lower limits respectively are given for the polarisation fraction. The dimensions and position angle of the synthesised beam in the image at each time are shown in the final column.}
    \label{KPNO-Tau_14_flux_density_table}
\end{table*}

\begin{table*}[h]
    \caption{Flux densities of KPNO-Tau 14 in Stokes I and V that were used in the light curve of the first burst in Fig. \ref{KPNO-Tau_14_1st_flare_light_curve}.}
    \centering
    \begin{tabular}{c c c c c}
    \hline 
    Time & Stokes I & Stokes V & Pol. Fraction & Beam \\
    (hrs) & (mJy) & (mJy) & (\%) &  \\
    \hline 
    1.50 - 1.55 & $\leq 8.40$ & $\leq 8.07$ & - & $18\farcs3 \times 3\farcs0$, $-60\degr$ \\ 
    1.55 - 1.60 & $\leq 8.76$ & $\leq 9.21$ & - & $18\farcs5 \times 3\farcs0$, $-60\degr$ \\ 
    1.60 - 1.65 & $13.76 \pm 2.77$ & $\leq 7.98$ & $\leq 58$ & $18\farcs2 \times 3\farcs0$, $-61\degr$ \\ 
    1.65 - 1.70& $17.81 \pm 2.90$ & $18.30 \pm 2.61$ & $88_{-13}^{+8}$ & $18\farcs1 \times 3\farcs0$, $-61\degr$ \\ 
    1.70 - 1.75 & $28.15 \pm 3.02$ & $27.10 \pm 2.58$ & $90_{-8}^{+7}$ & $18\farcs0 \times 3\farcs0$, $-62\degr$ \\ 
    1.75 - 1.80 & $20.67 \pm 2.81$ & $24.30 \pm 2.58$ & $93_{-8}^{+5}$ & $17\farcs9 \times 3\farcs0$, $-62\degr$ \\ 
    1.80 - 1.85 & $11.06 \pm 2.78$ & $17.10 \pm 2.52$ & $92_{-10}^{+6}$ & $17\farcs9 \times 3\farcs0$, $-63\degr$ \\ 
    1.85 - 1.90 & $\leq 7.86$ & $\leq 7.62$ & - & $17\farcs5 \times 3\farcs0$, $-63\degr$ \\ 
    1.90 - 1.95 & $\leq 7.89$ & $\leq 7.95$ & - & $16\farcs6 \times 3\farcs2$, $-63\degr$ \\ 
    1.95 - 2.00 & $\leq 9.42$ & $\leq 8.48$ & - & $16\farcs8 \times 3\farcs1$, $-64\degr$ \\ 

    \hline
    \end{tabular}
    \tablefoot{The time in the first column refers to the time in hours after the start of the observation (2013-11-30 19:15:00). The polarisation fraction is shown for the times where both Stokes I and V are detected. When only Stokes I or V are detected, $3\sigma$ upper or lower limits respectively are given for the polarisation fraction. The dimensions and position angle of the synthesised beam in the image at each time are shown in the final column.}
    \label{KPNO-Tau_14_burst_1_flux_density_table}
\end{table*}

\begin{table*}[h]
    \caption{Flux densities of LkCa 4 in Stokes I and V that were used in the light curve in Fig. \ref{LkCa_4_light_curve}.}
    \centering
    \begin{tabular}{c c c c c}
    \hline 
    Time & Stokes I & Stokes V & Pol. Fraction & Beam \\ 
    (hrs) & (mJy) & (mJy) & (\%) &  \\
    \hline 
    0 - 0.5 & $\leq 4.38$ & $\leq 3.21$ & - & $26\farcs8 \times 3\farcs5$, $-46\degr$ \\ 
    0.5 - 1.0 & $\leq 3.24$ & $\leq 2.61$ & - & $25\farcs1 \times 3\farcs5$, $-50\degr$ \\ 
    1.0 - 1.5 & $4.99 \pm 0.99$ & $3.30 \pm 0.70$ & $64_{-16}^{+18}$ & $23\farcs8 \times 3\farcs3$, $-55\degr$ \\ 
    1.5 - 2.0 & $4.30 \pm 0.76$ & $3.76 \pm 0.61$ & $80_{-15}^{+13}$ & $22\farcs7 \times 3\farcs1$, $-59\degr$ \\ 
    2.0 - 2.5 & $2.76 \pm 0.68$ & $3.00 \pm 0.56$ & $84_{-16}^{+11} $ & $21\farcs7 \times 2\farcs9$, $-65\degr$ \\ 
    2.5 - 3.0 & $\leq 1.89$ & $\leq 1.47$ & - & $20\farcs8 \times 2\farcs9$, $-71\degr$ \\ 
    3.0 - 3.5 & $\leq 2.01$ & $\leq 1.47$ & - & $19\farcs6 \times 3\farcs0$, $-77\degr$ \\ 
    3.5 - 4.0 & $\leq 1.89$ & $\leq 1.44$ & - & $15\farcs2 \times 4\farcs9$, $66\degr$ \\ 
    4.0 - 4.5 & $\leq 1.80$ & $\leq 1.50$ & - & $21\farcs2 \times 3\farcs8$, $84\degr$ \\ 
    4.5 - 5.0 & $\leq 1.74$ & $\leq 1.44$ & - & $13\farcs4 \times 4\farcs0$, $67\degr$ \\ 
    5.0 - 5.5 & $\leq 1.74$ & $\leq 1.47$ & - & $6\farcs2 \times 2\farcs3$, $-52\degr$ \\ 
    5.5 - 6.0 & $2.28 \pm 0.62$ & $\leq 1.53$ & $\leq 67$ & $9\farcs8 \times 3\farcs9$, $62\degr$ \\ 
    6.0 - 6.5 & $\leq 1.83$ & $\leq 1.53$ & - & $7\farcs5 \times 3\farcs9$, $61\degr$ \\ 
    6.5 - 7.0 & $3.95 \pm 0.60$ & $2.40 \pm 0.53$ & $60_{-15}^{+17}$ & $6\farcs5 \times 4\farcs0$, $60\degr$ \\ 
    7.0 - 7.5 & $4.46 \pm 0.65$ & $3.90 \pm 0.56$ & $82_{-13}^{+11}$ & $6\farcs1 \times 4\farcs0$, $59\degr$ \\ 
    7.5 - 8.0 & $\leq 2.25$ & $\leq 1.86$ & - & $6\farcs0 \times 4\farcs0$, $57\degr$ \\ 
    \hline 
    \end{tabular}
    \tablefoot{The time in the first column refers to the time in hours after the start of the observation (2019-08-10 02:11:00). The polarisation fraction is shown for the times where both Stokes I and V are detected. When only Stokes I is detected, $3\sigma$ upper limits are given for the polarisation fraction. The dimensions and position angle of the synthesised beam in the image at each time are shown in the final column.}
    \label{LkCa_4_flux_density_table}
\end{table*}

\end{appendix}
\end{document}